\documentstyle[12pt]{article}
\begin{document}
\title{\begin{flushright}
\small HUB-EP-98/27
\end{flushright}\medskip
\bf Relativistic properties of the quark-antiquark potential}
\author{ D. Ebert and R. N. Faustov\thanks{On leave of absence from
Russian Academy of Sciences, Scientific Council for
Cybernetics, Vavilov Street 40, Moscow 117333, Russia} \\
\small\it Institut f\"ur Physik, Humboldt--Universit\"at zu Berlin,\\
\small\it Invalidenstr.110, D-10115 Berlin, Germany\\
\\
 V. O. Galkin \\
\small\it Russian Academy of Sciences, Scientific Council for
Cybernetics,\\
\small\it Vavilov Street 40, Moscow 117333, Russia}
\date{}
\maketitle
\begin{abstract}
The relativistic transformation properties of the heavy  quark-antiquark
interaction potential are considered in the framework of the 
relativistic quark model. A special attention is paid to the
long-range (confining) contribution to the spin-independent part 
of $q\bar q$ interaction. The retardation effects are consistently
taken into account.
\end{abstract}

The relativistic properties of the quark-antiquark interaction potential
play an important role in analysing different static and dynamical 
characteristics of heavy me\-sons. The Lorentz-structure of the confining 
quark-anti\-quark interaction is of particular interest. In the literature 
there is no consent on this item. For a 
long time the scalar confining kernel has been considered to be the
most appropriate one \cite{s,scal}. The main argument in favour of this
choice is based on the nature of the heavy quark spin-orbit potential.
The scalar potential gives a vanishing long-range magnetic 
contribution, which is in agreement with the flux tube picture
of quark confinement of~\cite{b}, and allows to get the fine
structure for heavy quarkonia in accord with experimental data. 
However, the calculations of electroweak decay rates of heavy mesons 
with a scalar confining potential alone yield results which are in  worse 
agreement with data than for a vector potential \cite{mb,gf}. 
The  radiative
$M1$-transitions in quarkonia such as e.~g. $J/\psi\to \eta_c
\gamma$ are the most sensitive
for the Lorentz-structure of the confining potential. 
The relativistic corrections for 
these decays arising from vector and scalar potentials have different
signs \cite{mb,gf}. In particular, as it has been 
shown in ref.~\cite{gf}, agreement
with experiments for these decays can be achieved only for a mixture
of vector and scalar potentials. In this context, it is worth remarking,
that the recent study of the $q\bar q$ 
interaction in the Wilson loop approach \cite{bv1} indicates that
it cannot be considered as simply a scalar. Moreover, the found
structure of spin-independent relativistic corrections is not 
compatible with a scalar potential. A similar conclusion
has been obtained in ref.~\cite{ss} on the basis of a Foldy-Wouthuysen 
reduction of the full Coulomb gauge Hamiltonian of QCD. There, the 
Lorentz-structure of confinement has been found to be of vector
nature. The scalar character of spin splittings in heavy quarkonia
in this approach is dynamically generated through the interaction
with collective gluonic degrees of freedom. Thus we see that the
spin-dependent structure of ($q\bar q$) interaction is well established
now, while the spin-in\-de\-pen\-dent part is still controversial in the 
literature.

In preceding papers \cite{gmf,fg} we have developed the relativistic
quark model with the ($q\bar q$) potential consisting of the
perturbative one-gluon exchange part and a nonperturbative one which
is a mixture of the Lorentz scalar and vector confining potentials:
\begin{eqnarray}
\label{qpot}
V({\bf p,q};M)&=&\bar{u}_a({\bf p})
\bar{u}_b({\bf-p})\Bigg\{\frac{4}{3}\alpha_sD_{ \mu\nu}({
k})\gamma_a^{\mu}\gamma_b^{\nu}\cr
& & +V^V_{\rm conf}({k})\Gamma_a^{\mu}
\Gamma_{b;\mu}+V^S_{\rm conf}({
k})\Bigg\}u_a({\bf q})u_b({\bf-q}),
\end{eqnarray}
where $k=p-q$,  $D_{\mu\nu}$ is the gluon propagator in the Coulomb gauge
and $\Gamma_{\mu}$ is the effective vector long-range vertex, containing
both the Dirac and Pauli terms
\begin{equation}
\label{vert}
\Gamma_{\mu}=\gamma_{\mu}+
\frac{i\kappa}{2m}\sigma_{\mu\nu}k^{\nu},
\end{equation}
$u_{a,b}({\bf p})$ are the Dirac bispinors. Using the identity
$$\bar u({\bf p})\Gamma_{\mu} u({\bf q})=\bar u({\bf p})\left\{
\frac{p_{\mu}+q_{\mu}}{2m}+\frac{i(1+\kappa)}{2m}\sigma_{\mu\nu}k^{\nu}
\right\} u({\bf q})$$
we can treat the parameter $(1+\kappa)$ as the nonperturbative
(long-range) chromomagnetic moment of the quark and $\kappa$ as its
anomalous part (flavour independent).

In the nonrelativistic limit the Fourier transform of eq.~(\ref{qpot})
gives the static potential
\begin{equation}
\label{stat}
V_0(r)= V_{\rm Coul}(r)+V^S_{\rm conf}(r)+V^V_{\rm conf}(r),
\end{equation}
where
$$V_{\rm Coul}(r)=-\frac43\frac{\alpha_s}{r}.$$

In order to reproduce the linear confining potential
$$V_{\rm conf}(r)=Ar+B$$
in this limit one should put
\begin{eqnarray}
\label{lin}
\nonumber V^V_{\rm conf}(r)&=&(1-\varepsilon)(Ar+B)\\
V^S_{\rm conf}(r)&=&\varepsilon (Ar +B),
\end{eqnarray}
where $\varepsilon$ is the mixing parameter.

Now assuming that both quarks are heavy enough we evaluate the $(v^2/c^2)$
relativistic corrections to the static potential (\ref{stat}), (\ref{lin}).
For the one-gluon exchange part the retardation effect is taken into
account by the contribution of transverse gluon exchange and thus it is
sufficient in the adopted approximation to set $k^0=0$. For the confining
part one should utilize a different procedure (see \cite{ab,nf,g}).
The Fourier transform of the linear potential $Ar$ in
the momentum space looks like:
\begin{equation}
\label{four}
A\int {\rm d}^3 r re^{-i{\bf k\cdot r}}=-A\frac{8\pi}{|{\bf k}|^4},
\quad {\bf k}={\bf p}-{\bf q}.
\end{equation}
The natural (though not unique) relativistic extension (dependent
only on the four-momentum transfer) of expression (\ref{four}) is to
substitute $(-{\bf k}^2)\to (k_0^2-{\bf k}^2)$ and thus
\begin{equation}
\label{k}
\frac{1}{|{\bf k}|^4} \to \frac{1}{(k_0^2-{\bf k}^2)^2}.
\end{equation}
Now as mentioned above we should choose the procedure of fixing $k_0$.
On the mass shell due to energy conservation we have  $k_0=0$. So $k_0$
may be considered as the measure of deviation either from  the mass shell
or from the energy shell. We choose the second possibility and set  $k_0$
equal to $\epsilon_a({\bf p}) -\epsilon_a({\bf q})$ or to $\epsilon_b({\bf
q})-\epsilon_b({\bf p})$. Then in the symmetrized form \cite{ab,g} we get
\begin{eqnarray} \label{ko}
&&k_0^2=-(\epsilon_a({\bf p})-\epsilon_a({\bf q}))(\epsilon_b({\bf p})
-\epsilon_b({\bf q})),\\
&&\epsilon_{a,b}({\bf p})=\sqrt{{\bf p}^2+m_{a,b}^2}\nonumber.
\end{eqnarray}
This form is not unique and other possible expressions for $k_0^2$ are
discussed in \cite{g,om}. In favour of choice (\ref{ko}) we mention the
following arguments. It is well-known \cite{ab,ch} that for the one-photon
exchange contribution in QED only choice (\ref{ko}) in the Feynman
(diagonal) gauge leads to the same correct result (the Breit-Fermi
Hamiltonian) as the prescription $k_0=0$ in the Coulomb (or transverse
Landau) gauge. The same is naturally true for the one-gluon exchange
contribution in QCD. Moreover as shown in ref.~\cite{ch} for any
effective vector potential generated by a vector exchange and its
couplings to conserved vector currents (vertices) there is the so-called
instantaneous gauge which plays the role of the Coulomb gauge. In the
instantaneous gauge the prescription $k_0=0$ reproduces the same result
as the expansion in $k_0^2$ fixed by eq.~(\ref{ko}) in the diagonal
gauge used here. The other reason to utilize prescription (\ref{ko})
is the reproduction of the correct Dirac limit in this case \cite{om}.

The $(p^2/m^2)$ expansion of (\ref{k}) with the account of
eq.~(\ref{ko}) yields:
\begin{eqnarray}
\label{rk}
&&\frac{1}{[(\epsilon_a({\bf p})-\epsilon_a({\bf q}))(\epsilon_b({\bf p})
-\epsilon_b({\bf q}))
+{\bf k}^2]^2}
\cong \frac{1}{|{\bf k}|^4}\left[1-
\frac{({\bf p}^2-{\bf q}^2)^2}{2m_am_b|{\bf k}|^2}\right]\cr 
&&\qquad\qquad= \frac{1}{|{\bf k}|^4} -\frac{1}{2m_am_b|{\bf k}|^6}\big[
({\bf k\cdot p})^2
+2({\bf k\cdot p})({\bf k\cdot q})+({\bf k\cdot q})^2
\big].
\end{eqnarray}
After the Fourier transform of eq.~(\ref{rk}) we obtain in the
configuration space:
\begin{eqnarray}
\label{ar}
&&-8\pi A\int\frac{{\rm d}^3k}{(2\pi)^3} e^{i{\bf k\cdot r}}
\frac{1}{|{\bf k}|^4}\left[1-
\frac{({\bf p}^2-{\bf q}^2)^2}{2m_am_b|{\bf k}|^2}\right] 
=Ar-\left\{\frac{Ar}{2m_am_b}\left[{\bf p}^2+\frac{({\bf p\cdot r})^2}
{r^2}\right]\right\}_W,
\end{eqnarray}
where the notation $\{\dots\}_W$ means the Weyl ordering prescription for
${\bf p}$ and ${\bf r}$ variables.

Now we turn to the constant $B$ term in eq.~(\ref{lin}). The Fourier
transform of it gives in the momentum space
\begin{equation}
\label{const}
B\int {\rm d}^3r e^{-i{\bf k\cdot r}}= B(2\pi)^3\delta({\bf k}),
\quad {\bf k}={\bf p}-{\bf q}.
\end{equation}
The simplest relativistic version of eq.~(\ref{const}) is multiplying it
by the energy factor $\epsilon({\bf p})/m$, which in the symmetric form
looks like
\begin{equation}
\label{delt}
\sqrt{\frac{\epsilon_a({\bf p})\epsilon_b({\bf p})}{m_am_b}}\delta({\bf
p-q}).
\end{equation}
Expanding eq.~(\ref{delt}) in $(p^2/m^2)$ we get
\begin{equation}
\label{edelt}
\delta({\bf p-q})\left[1+\frac{{\bf p}^2}{4}\left(\frac{1}{m_a^2}+
\frac{1}{m_b^2}\right)\right].
\end{equation}
So in the configuration space the constant term acquires the form
\begin{equation}
\label{econst}
B\left[1+\frac14\left(\frac{1}{m_a^2}+\frac{1}{m_b^2}\right){\bf p}^2
\right].
\end{equation}

All other relativistic corrections of order $(p^2/m^2)$ have been
considered in ref.~\cite{gmf}. After using the Weyl ordering notations, 
expressions (9) of ref.~\cite{gmf} take on the form: 

\begin{eqnarray}
&&V_0(r) +\frac18\left(\frac{1}{m_a^2}+\frac{1}{m_b^2}\right)
\Delta V_{\rm Coul}(r)
+\frac18\left(\frac{1+2\kappa_a}{m_a^2}+
\frac{1+2\kappa_b}{m_b^2}\right)\Delta V^V_{\rm
conf}(r)\cr
&&+\frac{1}{2m_am_b}\left\{V_{\rm Coul}\left[{\bf p}^2+
\frac{({\bf p\cdot r})^2}{r^2}\right]\right\}_W \cr
&& +\frac{1}{m_am_b}\left\{V^V_{\rm conf}(r){\bf p}^2\right\}_W
-\frac{1}{2}\left(\frac{1}{m_a^2}
+\frac{1}{m_b^2}\right)
\left\{V^S_{\rm conf}(r){\bf p}^2\right\}_W. \nonumber
\end{eqnarray}
Adding contributions (\ref{ar}) and (\ref{econst}) to the above expressions
we obtain
the complete spin-independent  part of the $(q\bar q)$
potential ($\kappa_a=\kappa_b=\kappa$):
\begin{eqnarray}
\label{spind}
&&V_{\rm SI}(r)=V_0(r) + V_{\rm VD}(r)+\frac18\left(\frac{1}{m_a^2}+
\frac{1}{m_b^2}\right)
\Delta\big[V_{\rm Coul}(r) 
 +(1+2\kappa)V^V_{\rm
conf}(r)\big],
\end{eqnarray}
where $V_0(r)$ is given by eqs.~(\ref{stat}), (\ref{lin}).
For the velocity-dependent part $V_{\rm VD}(r)$ we have
\begin{eqnarray}
\label{vd}
V_{\rm VD}(r)&=&\frac{1}{2m_am_b}\left\{\left(-\frac43\frac{\alpha_s}{r}
\right)\left[{\bf p}^2+\frac{({\bf p\cdot r})^2}{r^2}\right]\right\}_W \cr
& & +\frac{(1-\varepsilon)}{2m_am_b}\left\{Ar\left[{\bf p}^2-
\frac{({\bf p\cdot r})^2}{r^2}\right]\right\}_W \cr
& &-\frac{\varepsilon}{2}\left(\frac{1}{m_a^2}+\frac{1}{m_b^2}\right)
\left\{Ar{\bf p}^2\right\}_W
-\frac{\varepsilon}
{2m_am_b}\left\{Ar\left[{\bf p}^2+\frac{({\bf p\cdot r})^2}{r^2}\right]
\right\}_W \cr
&&-\left(\frac{\varepsilon}{2}-\frac14\right)
\left(\frac{1}{m_a^2}+\frac{1}{m_b^2}\right)B{\bf p}^2
+\frac{1-\varepsilon}{m_am_b}B{\bf p}^2.
\end{eqnarray}
Now representing eq.~(\ref{vd}) in the form
\begin{eqnarray}
\label{vdrel}
&&V_{\rm VD}(r)=\frac{1}{m_am_b}\left\{{\bf p}^2V_{bc}(r)+\frac{({\bf p
\cdot r})^2}{r^2}V_c(r)\right\}_W \cr
& &\qquad+\left(\frac{1}{m_a^2}+\frac{1}{m_b^2}\right)\left\{{\bf p}^2 V_{de}
(r) -\frac{({\bf p\cdot r})^2}{r^2}V_e(r)\right\}_W
\end{eqnarray}
with
\begin{eqnarray}
\label{coef}
&&V_{bc}(r)=-\frac{2\alpha_s}{3r}+\left(\frac12-\varepsilon\right)Ar
+(1-\varepsilon)B;\cr
&&V_{de}(r)=-\frac{\varepsilon}{2}Ar+\left(\frac14-\frac{\varepsilon}{2}
\right) B;\cr
&&V_c(r)=-\frac{2\alpha_s}{3r}-\frac12Ar; \quad V_e(r)=0,
\end{eqnarray}
we are able to test the fulfilment of the exact Barchielli, Brambilla,
Prosperi (BBP)
relations \cite{BBP}, which follow from the Lorentz invariance
of the Wilson loop. In our notations these relations look like
\begin{eqnarray}
\label{re}
&&V_{de}-\frac12 V_{bc}+\frac14V_0=0 \cr
&&V_e+\frac12 V_c+\frac{r}{4}\frac{{\rm d} V_0}{{\rm d} r}=0
\end{eqnarray}
(in the original version $V_{bc}\equiv-V_b-\frac13 V_c$ and 
$V_{de}\equiv V_d+\frac13
V_e$). One can easily find that the functions (\ref{coef}) identically
satisfy relations (\ref{re}) independently of values of the parameters
$\varepsilon$ and $\kappa$. This is a highly nontrivial result. For
the perturbative one-gluon-exchange part of $V_{\rm VD}$ our expressions
for $V_b$, $\dots$, $V_e$ are the same as in \cite{BBP,bcp}, but for
the confining (long-range) part they are different, namely the result
of refs.~\cite{BBP,bcp} (from the minimal area law) is as follows:
\begin{eqnarray}
&& V_{bc}(r)=-\frac{2\alpha_s}{3r}+\frac16 Ar;
\quad V_c(r)=-\frac{2\alpha_s}{3r}-\frac16 Ar;\cr
&& V_{de}(r)=-\frac16 Ar-\frac14 B; \quad V_e=-\frac16 Ar.
\end{eqnarray}
No value of $\varepsilon$ in eqs.~(\ref{coef}) can reproduce the above
result. The terms with the Laplacian in (\ref{spind})
coincide only for $\kappa=0$ and $\varepsilon=0$, i.~e. for
purely vector confining interaction without the Pauli term in the vertex
(\ref{vert}). Our expressions (\ref{spind}) and (\ref{vd}) for purely
vector ($\varepsilon=0$) and purely scalar ($\varepsilon=1$)
interactions and for $\kappa=0$ coincide with those of ref.~\cite{om}
except for the constant $B$ term. Our $B$ term for $\varepsilon=1$
(scalar potential) is the same as in \cite{BBP}. The $B$ term from
ref.~\cite{om} does not satisfy the BBP relations (it gives
contribution $-B/2$ only to $V_{de}$). Our
result (\ref{vd}) for the scalar ($\varepsilon=1$) confining potential
also differs from the one obtained in ref.~\cite{bg}, where the
prescription $k_0=0$ was used and as a result the contribution of
retardation was lost. The differences between our results and the
results presented in ref.~\cite{bv} originate from the use of specific
models such as minimal area law, flux tube, dual superconductivity
and stochastic vacuum.

The spin-dependent part of the ($q\bar q$) potential is given in
ref.~\cite{gmf} ($\kappa_a=\kappa_b=\kappa$):
\begin{eqnarray}
\label{sd}
V_{\rm SD}(r)&=&\frac{1}{m_am_b}\frac{1}{r}\frac{\rm d}{{\rm d} r}
V_{\rm Coul}(r){\bf L}\cdot ({\bf s}_a+{\bf s}_b) \cr
& &+\frac{1}{2m_a^2}\frac1r\frac{\rm d}{{\rm d} r}\bigg\{[V_{\rm
Coul}(r)-V_{\rm conf}(r)]
+2(1+\kappa)\left(1+\frac{m_a}{m_b}\right)
V^V_{\rm conf}(r)\bigg\}{\bf L\cdot s}_a \cr
& & +\frac{1}{2m_b^2}\frac1r\frac{\rm d}{{\rm d} r}\bigg\{[V_{\rm
Coul}(r)-V_{\rm conf}(r)]
+2(1+\kappa)\left(1+\frac{m_b}{m_a}\right)
V^V_{\rm conf}(r)\bigg\}{\bf L\cdot s}_b \cr
& & + \frac{1}{3m_am_b}\bigg\{\frac1r\frac{\rm d}{{\rm d} r}[
V_{\rm Coul}(r) +(1+\kappa)^2V^V_{\rm conf}(r)] \cr
&&- \frac{{\rm d}^2}{{\rm d} r^2}[V_{\rm Coul}+(1+\kappa)^2V^V_{\rm
conf}(r)]\bigg\}
\left[\frac{3}{r^2}({\bf s}_a\cdot {\bf r})({\bf s}_a
\cdot {\bf r})-{\bf s}_a\cdot {\bf s}_b\right] \cr
&&+\frac{2}{3m_am_b}\Delta[V_{\rm Coul}(r) +(1+\kappa)^2V^V_{\rm conf}
(r)]{\bf s}_a\cdot {\bf s}_b, \quad
\end{eqnarray}
where ${\bf L}$ is the orbital momentum and ${\bf s}_{a,b}$ are the
spin momenta.

The correct description of the fine structure of the heavy quarkonium
mass spectrum requires vanishing of the vector confinement contribution.
This can be achieved by putting $1+\kappa=0$, i.e. the total
long-range quark chromomagnetic moment equals to zero, which is in accord
with the flux tube \cite{b} and minimal area \cite{bcp,bv} models.
One can see from eq.~(\ref{sd}) that for the spin-dependent part of
the potential this conjecture is equivalet to
the assumption about the scalar structure of confinement interaction
\cite{s}. The specific
value of vector-scalar mixing parameter $\varepsilon=-1$
provides the correct description of radiative decays of
heavy quarkonia \cite{gf}.

In this way setting $\kappa=-1$, we obtain:
\begin{eqnarray}
\label{vsd}
V_{\rm SD}(r)&=&\frac12\left(\frac{4\alpha_s}{3r^3}-\frac{A}{r}\right)
\left[\frac{1}{m_a^2}{\bf L\cdot s}_a+\frac{1}{m_b^2}{\bf L\cdot s}_b
\right] \cr
&&+\frac{1}{m_am_b}\left(\frac{4\alpha_s}{3r^3}\right)
{\bf L}\cdot ({\bf s}_a+
{\bf s}_b)
+\frac{8\pi}{3m_am_b}\left(\frac{4\alpha_s}{3}\right)
\delta({\bf r}){\bf s}_a\cdot{\bf s}_b \cr
&&+\frac{1}{m_am_b}\left(\frac{4\alpha_s}{3r^3}
\right)\left[\frac{3}{r^2}
({\bf s}_a\cdot {\bf r})({\bf s}_b\cdot {\bf r})-{\bf s}_a\cdot {\bf s}_b
\right]. \quad
\end{eqnarray}
Expression (\ref{vsd}) for $V_{\rm SD}$ completely coincides with the one
found in refs.~\cite{bcp,b}. The Gromes relation is identically
fulfilled. Our result supports the conjecture that the long-range
confining forces are dominated by chromoelectric interaction and
that the chromomagnetic interaction vanishes. It is also in
accord with the dual superconductivity picture \cite{bbbpz,bv}. It is 
important to mention, that our relativistic quark model is in complete
agreement with the heavy quark effective theory (HQET). The model
correctly reproduces  HQET results for heavy-to-heavy
and heavy-to-light weak transition
matrix elements with the specific choice of the parameters
$\varepsilon=-1$, $\kappa=-1$ (see ref.~\cite{fg} for details)
in accord with the ones found previously
\cite{gf,gmf,efg}. It includes the proper description of  invariant
form factors within the inverse heavy-quark-mass expansion up to the
terms of order of $1/m_Q^2$. The mass spectra of $D$ and $B$ mesons
have been also calculated in our model in complete agreement with the
HQET predictions and available experimental data \cite{egf}. It is
interesting to note that the relations which are equivalent to the
BBP relations (\ref{re}) can be obtained by use of the reparametrization
invariance (in four-velocity) within HQET \cite{ck}. The
phenomenological implications of retardation corrections will be
considered elsewhere.

We are grateful to D.V. Antonov, K.-J. Biebl,
N. Brambilla, G.M. Prosperi and V.I. Savrin for useful
discussions of the results. One of the authors (R.N.F.) would like to
thank the particle theory group of Humboldt University for the kind 
hospitality.
The work of R.N.F. and V.O.G. was supported
in part by the Deutsche Forschungsgemeinschaft under contract Eb 139/1-3
and in part by the Russian Foundation for Fundamental Research under
Grant No. 96-02-17171.

\end{document}